\documentclass{amsart}
\usepackage{amsmath,amssymb,amsthm}
\usepackage[longnamesfirst]{natbib}
\usepackage{setspace}
\usepackage{graphicx}
\usepackage{enumerate}         % better lists
\newtheorem{thm}{Theorem}[section]
\newtheorem{cor}[thm]{Corollary}
\newtheorem{lem}[thm]{Lemma}
\newtheorem{prop}[thm]{Proposition}
\theoremstyle{definition}
\newtheorem{defn}[thm]{Definition}

\theoremstyle{remark}
\newtheorem{rem}[thm]{Remark}

\numberwithin{equation}{section}

\newcommand{\Xt}{(X_t)_{t\geq0}}
\newcommand{\St}{(S_t)_{t\geq0}}
\newcommand{\Jt}{(J_t)_{t\geq0}}
\newcommand{\Vt}{(V_t)_{t\geq0}}

\newcommand{\set}[1]{\left\{#1\right\}}
\newcommand{\Ind}[1]{\mathbf{1}_{\left\{#1\right\}}}
\newcommand{\RR}{\mathbb{R}}

\newcommand{\CC}{\mathbb{C}}

\newcommand{\EE}{\mathbb{E}}
\newcommand{\cU}{\mathcal{U}}

\newcommand{\cF}{\mathcal{F}}

\newcommand{\Rplus}{\mathbb{R}_{\geqslant 0}}

\newcommand{\pd}[2]{\frac{\partial #1}{\partial #2}}

\newcommand{\uw}{\left(\begin{array}{@{}c@{}}u\\w\end{array}\right)}
\newcommand{\scal}[2]{\left\langle{#1},{#2}\right\rangle}

\renewcommand{\Re}{\mathrm{Re}}

\newcommand{\dom}{\mathrm{dom}}

\title[Affine Stochastic Volatility Models]{Moment Explosions and Long-Term Behavior of Affine Stochastic Volatility Models}
\author[Martin Keller-Ressel]{Martin Keller-Ressel}
\thanks{Supported by the Austrian Science Fund (FWF) through the START programm Y328.}
\address{Vienna University of Technology, Wiedner Hauptstrasse 8--10,
A-1040 Wien, Austria} \email{mkeller@fam.tuwien.ac.at}

\begin{document}
%\onehalfspace \doublespace

\begin{abstract} We consider a class of asset pricing models, where the risk-neutral joint process of log-price and its
stochastic variance is an affine process in the sense of
\citet*{Schachermayer}. First we obtain conditions for the price
process to be conservative and a martingale. Then we present some
results on the long-term behavior of the model, including an
expression for the invariant distribution of the stochastic
variance process. We study moment explosions of the price process,
and provide explicit expressions for the time at which a moment of
given order becomes infinite. We discuss applications of these
results, in particular to the asymptotics of the implied
volatility smile, and conclude with some calculations for the
Heston model, a model of Bates and the Barndorff-Nielsen-Shephard
model.
\end{abstract}

\maketitle

%\tableofcontents

\section{Introduction}\label{Sec:Introduction}

\citet*{DuffiePanSingleton} introduced the notion of an affine
jump-diffusion, which is a jump-diffusion process, whose drift
vector, instantaneous covariance matrix and arrival rate of jumps
all depend in an affine way on the state vector.
\citeauthor*{DuffiePanSingleton} remark that models built on
affine processes provide a balanced tradeoff between analytical
tractability and complexity, making them an attractive choice for
applications in mathematical finance. In particular they mention
applications to the pricing of options in stochastic volatility
models and note that the models of \citet{Heston},
\citet{Bates1996, Bates}
%\footnote{The article later published as
%\citet{Bates} is cited by \citet{DuffiePanSingleton} as a working
%paper of the National Bureau of Economic Research from 1997.}
, and
\citet{BakshiCaoChen} fall into the affine class. To these, we
could add the more recent models of \citet{BNS} and
\citet{CarrWu}, which are also of affine type.\\
\citet*{Schachermayer} subsequently extended the class of affine
jump-diffusions, defining an \emph{affine process} as a
time-homogenous Markov process, whose characteristic function is
the exponential of an affine function of the state vector. It
turns out that this class coincides for a large part with the
class of affine jump-diffusions, but also allows for infinite
activity of jumps and for killing or explosions of the process.
\citeauthor*{Schachermayer} aim to give a rigorous mathematical
foundation to the theory of affine processes, covering many
aspects, such as the characterization of an affine process in
terms of the `admissible parameters' (comparable to the
\emph{characteristic triplet} of a L\'evy process) and properties
of the ordinary differential equations (`generalized Riccati
equations') that are implied by the process.\\

In this article we study stochastic volatility models, comprised
of a log-price process $\Xt$ and a stochastic variance process
$\Vt$, such that the joint process $(X_t,V_t)_{t \geq 0}$ is an
\emph{affine process.} We will show that many properties of such a
model, including its long-term behavior and moment explosions, can
be analyzed by studying differential equations of the generalized
Riccati type. Our results on the long-term behavior are formulated
as asymptotic results for the cumulant generating function of the
stock price, as time goes to infinity. Asymptotics of this type
have been used by \citet{Lewis} to obtain large-time-to-maturity
results for the implied volatility smile of stochastic volatility
models via a saddlepoint expansion. The issue of moment explosions
in stochastic volatility models has recently received much
attention, due to the articles of \citet{Piterbarg} and
\citet{LionsMusiela}. Moment explosions are intimately connected
to large-strike asymptotics of the implied volatility smile via
results of \citet{Lee}, that have later been expanded by
\citet{Friz}.\\

In the first part of the paper we introduce our main assumption,
that the joint process $(X_t,V_t)_{t \geq 0}$ is affine, and
recapitulate the main results of \citet{Schachermayer}. We derive
necessary and sufficient conditions for conservativeness of the
process and for the martingale property of the discounted price
process $S_t = \exp(X_t)$. At the end of Section~\ref{Sec:ASVM} we
add two assumptions, and give a precise definition of the class of
affine stochastic volatility models, which constitutes the main
subject of this article. In Section~\ref{Sec:LongTerm} we derive
our central results on long-term properties of an affine
stochastic volatility model, providing conditions for the
existence of an invariant distribution of the stochastic variance
process, and characterizing this distribution in terms of its
cumulant generating function. We also give results on the
long-term properties of the price process, showing that as time
tends to infinity, the marginal distributions of the price process
approach those of an exponential-L\'evy process. The
characteristic exponent of this L\'evy process can be derived
directly from the specification of the affine stochastic
volatility model. Both results are obtained by applying
qualitative ODE theory to the generalized Riccati
equations introduced in the first part.\\
In Section~\ref{Sec:MomentExplosion} we study moment explosions of
the price process, and show that an explicit representation for
the time of moment explosion can be given -- not only for the
primary model, but also for the model in the stationary variance
regime. In Section~\ref{Sec:Applications} we outline applications
of our results to the asymptotics of implied volatilities and of
implied forward volatilities. We briefly discuss the results of
\citet{Lee} and point out the connection between the stationary
variance regime and the pricing of forward-start options, when the
time until the start of the contract is large. We conclude in
Section~\ref{Sec:Examples} with explicit calculations for several
models to which our results apply, such as the Heston model, a
Heston model with added jumps, a model of \citeauthor{Bates}, and
the Barndorff-Nielsen-Shephard model.

\section{Affine Stochastic Volatility Models}\label{Sec:ASVM}
\subsection{Definition and the generalized Riccati equations}
We consider an asset-pricing model of the following kind: The
interest rate $r$ is non-negative and constant, and the asset
price $(S_t)_{t \geq 0}$ is given by
\[S_t = \exp(rt + X_t)\quad t \geq 0\;,\]
such that $\Xt$ is the discounted log-price process starting at
$X_0 \in \RR$ a.s. The discounted price process is simply
$\exp(X_t)$, such that we will assume in the remainder that $r =
0$, and that $\St$ is already discounted. Denote by $(V_t)_{t \geq
0}$ another process, starting at $V_0
> 0$ a.s., which can be interpreted as stochastic variance process
of $\Xt$, but may also control the arrival rate of jumps. The
following assumptions are made on the joint process $(X_t, V_t)_{t
\geq 0}$:
\begin{description}
\item[A1] $(X_t,V_t)_{t \geq 0}$ is a stochastically continuous,
time-homogeneous Markov process. \label{Eq:main_ass_sub1}
\item[A2] The cumulant generating function $\Phi_t(u,w)$ of
$(X_t,V_t)$ is of a particular affine form: We assume that there
exist functions $\phi(t,u,w)$ and $\psi(t,u,w)$ such that
\begin{equation*}\label{Eq:affine}
\Phi_t(u,w) := \log \EE\left[\left.\exp(u X_t + w
V_t)\right|X_0,V_0\right] = \phi(t,u,w) + V_0 \psi(t,u,w) + X_0 u
\end{equation*}
for all $(t,u,w) \in \Rplus \times \CC^2$, where the expectation
exists.\label{Eq:main_ass_sub2}
\end{description}

By convention, the logarithm above denotes the principal branch of
the complex logarithm. Assumptions A1 and A2 make $(X_t,V_t)_{t
\geq 0}$ an affine process in the sense of \cite{Schachermayer}.
The term $X_0 u$ in the cumulant generating function $\Phi_t(u,w)$
corresponds to a reasonable homogeneity assumption on the model:
If the starting value $X_0$ of the price process is shifted by
$x$, also $X_t$ is simply shifted by $x$ for any $t \ge 0$. Note
that Assumption A2 also implies that the variance process
$(V_t)_{t \geq 0}$ is a Markov process in its own right. We do not
yet make the assumption that $\St$ is conservative (i.e. without
explosions or killing) or even a martingale. Instead it will be
our first goal in Section~\ref{Sec:martingale} to obtain necessary
and sufficient conditions for these
properties.\\
Applying the law of iterated expectations to $\Phi_t(u,w)$ yields
the following `flow-equations' for $\phi$ and $\psi$: (see also
\citet[Eq.~(3.8)--(3.9)]{Schachermayer})
\begin{equation}\label{Eq:flow_prop}
\begin{split}
\phi(t+s,u,w) &= \phi(t,u,w) + \phi(s,u,\psi(t,u,w)),\\
\psi(t+s,u,w) &= \psi(s,u,\psi(t,u,w)),
\end{split}
\end{equation}
for all $t, s \ge 0$. The following result will be crucial:

\begin{thm}\label{Thm:Riccati}
Suppose that $|\phi(\tau,u,\eta)| < \infty$ and
$|\psi(\tau,u,\eta)| < \infty$ for some $(\tau,u,\eta) \in \Rplus
\times \CC^2$. Then, for all $t \in [0,\tau]$ and $w \in \CC$ with
$\Re\,w \le \Re\,\eta$
\[|\phi(t,u,w)| < \infty, \qquad |\phi(t,u,w)| < \infty,\]
and the derivatives
\begin{equation} F(u,w) :=
\left.\pd{}{t}\phi(t,u,w)\right|_{t = 0+}, \qquad R(u,w) :=
\left.\pd{}{t}\psi(t,u,w)\right|_{t = 0+}
\end{equation}
exist. Moreover, for $t \in [0,\tau)$, $\phi$ and $\psi$ satisfy
the generalized Riccati equations
\begin{subequations}\label{Eq:gen_Riccati}
\begin{align}
\partial_t \phi(t,u,w) &= F(u,\psi(t,u,w)), \quad \phi(0,u,w) = 0 \label{Eq:gen_Riccati_sub1}\\
\partial_t \psi(t,u,w) &= R(u,\psi(t,u,w)), \quad \psi(0,u,w) =
w\;.\label{Eq:gen_Riccati_sub2}
\end{align}
\end{subequations}
\end{thm}
The above theorem is `essentially' proven in
\citet{Schachermayer}, but under slightly different
conditions\footnote{\citeauthor{Schachermayer} assume
differentiability of $\phi$ and $\psi$ with respect to $t$
('regularity') a priori, while in our case we can deduce it
directly from Assumption A2. A proof is given in the appendix.}.
Note that the differential equations \eqref{Eq:gen_Riccati} follow
immediately from the flow equations \eqref{Eq:flow_prop} by taking
the derivative with respect to $s$, and evaluating at $s = 0$.
They are called \emph{generalized Riccati Equations} since they
degenerate into (classical) Riccati
equations with quadratic functions $F$ and $R$, if $(X_t, V_t)_{t \ge 0}$ is a pure diffusion process.\\
Note that the first Riccati equation is just an integral in
disguise, and $\phi$ may be written explicitly as
\begin{equation}\label{Eq:Riccati_explicit}
\phi(t,u,w) = \int_0^t{F(u,\psi(s,u,w))\;ds}\;.
\end{equation}
Also the solution $\psi$ of the second Riccati equation can be
represented at least implicitly in the following way: Suppose that
$\psi(t,u,w)$ is a non-stationary local solution on $[0,\delta)$
of \eqref{Eq:gen_Riccati_sub2}. Then $R(u,\psi(t,u,w)) \neq 0$ for
all $t \in [0,\delta)$, and $\psi(t,u,w)$ is a strictly monotone
function of $t$; dividing both sides of
\eqref{Eq:gen_Riccati_sub2} by $R(u,\psi(t,u,w))$, integrating
from $0$ to $t < \delta$, and substituting $\eta = \psi(s,u,w)$
yields
\begin{equation}\label{Eq:eta_integral}
    \int_w^{\psi(t,u,w)}{\frac{d\eta}{R(u,\eta)}ds} = t\;.
\end{equation}
%Often \eqref{Eq:eta_integral} can be inverted to give an explicit
%expression for $\psi$.\\

Another important result that can be found in
\citet{Schachermayer} states that $F$ and $R$ must be of
L\'evy-Khintchine form, i.e.
\begin{subequations}\label{Eq:FR_form}
\begin{align}
F(u,w) &= (u,w) \cdot \frac{a}{2} \cdot \uw + b \cdot \uw - c \\
&+\int_{D \setminus \set{0}}{\left(e^{xu + yw} - 1 - \omega_F(x,y) \cdot \uw \right)\,m(dx,dy)} \notag\;,\\
R(u,w) &= (u,w) \cdot \frac{\alpha}{2} \cdot \uw + \beta \cdot \uw
 - \gamma \\
 &+\int_{D \setminus \set{0}}{\left(e^{xu + yw} - 1 - \omega_R(x,y)
\cdot \uw \right)\,\mu(dx,dy)} \notag
\end{align}
\end{subequations}
where $D = \RR \times \Rplus$, and $\omega_F$, $\omega_R$ are
suitable truncation functions, which we fix by defining
\begin{equation*}
\omega_F(x,y) = \left(\begin{array}{@{}c@{}}\frac{x}{1+x^2}\\0
\end{array}\right) \qquad \text{and} \qquad
\omega_R(x,y) =
\left(\begin{array}{@{}c@{}}\frac{x}{1+x^2}\\\frac{y}{1+y^2}
\end{array}\right)\;.
\end{equation*}

Moreover the parameters $(a,\alpha,b,\beta,c,\gamma,m,\mu)$
satisfy the following admissibility conditions:
\begin{itemize}
\item $a, \alpha$ are positive semi-definite $2 \times
2$-matrices, and $a_{12} = a_{21} = a_{22} = 0$. \item $b \in D$
and $\beta \in \RR^2$. \item $c, \gamma \in \Rplus$ \item $m$ and
$\mu$ are L\'evy measures on $D$, and $\int_{D \setminus
\set{0}}{\left((x^2 + y) \wedge 1\right)\,m(dx,dy)} < \infty$.
\end{itemize}

%Again \eqref{Eq:FR_form} are obtained under slightly different
%conditions in \citet{Schachermayer} such that we also give a proof
%in the appendix.
The affine form of the cumulant generating function, the
generalized Riccati equations and finally the L\'evy-Khintchine
decomposition \eqref{Eq:FR_form} lead to the following
interpretation of $F$ and $R$: $F$ characterizes the
state-independent dynamic of the process $(X_t,V_t)$ while $R$
characterizes its state-dependent dynamic. Both $F$ and $R$
decompose into a diffusion part, a drift part, a jump part and an
instantaneous killing rate. Hence $a + \alpha V_t$ can be regarded
as instantaneous covariance matrix of $(X_t,V_t)_{t \geq 0}$, $b +
V_t \beta$ as the instantaneous drift, $m(dx,dy) + V_t \mu(dx,dy)$
as instantaneous arrival rate of jumps with jump heights in $(dx
\times dy)$, and finally $c + \gamma V_t$ as the instantaneous
killing rate.\\

The following Lemma establishes some important properties of $F$
and $R$ as functions of \emph{real-valued} arguments. A proof is
given in the appendix.
\begin{lem}\label{Lem:psiphi_prop}
\begin{enumerate}[(a)]
\item \label{Eq:psiphi_sub3} $F$ and $R$ are proper closed convex
functions on $\RR^2$. \item \label{Eq:psiphi_sub4}$F$ and $R$ are
analytic in the interior of their effective domain. \item Let $U$
be a one-dimensional affine subspace of $\RR^2$. Then $F|_U$ is
either a strictly convex or an affine function. The same holds for
$R|_U$. \item If $(u,w) \in \dom\,F$, then also $(u,\eta) \in
\dom\,F$ for all $\eta \le w$. The same holds for $R$.
\end{enumerate}
\end{lem}
\begin{rem}
As usual in convex analysis, we regard $F$ and $R$ as functions
defined on all of $\RR^2$, that may attain values in $\RR \cup
\set{+\infty}$. The set $\set{(u,w) : F(u,w) < \infty}$ is called
effective domain of $F$, and denoted by $\dom\,F$.\\
\end{rem}

We define a function $\chi(u)$, that will appear in several
conditions throughout this article.
Corollary~\ref{Cor:phipsi_estimate} gives an interpretation of
$\chi$ as a rate of convergence for the asymptotic behavior of the
cumulant generating function of $\Xt$.

\begin{defn}
For each $u \in \RR$ where $R(u,0) < \infty$, define $\chi(u)$ as
\[\chi(u) := \left.\pd{R}{w}(u,w)\right|_{w = 0}\;.\]
\end{defn}

$\chi(u)$ is well-defined at least as a limit as $w \uparrow 0$,
possibly taking the value $+\infty$; it can be written explicitly
as
\[\chi(u) = \alpha_{12} u + \beta_1 + \int_{D \setminus \set{0}}{y \left(e^{xu} - \frac{1}{1 + y^2}\right)\,\mu(dx,dy)}\;.\]
Note that also $\chi(u)$ is a convex function.

\subsection{Explosions and the martingale property}\label{Sec:martingale}

We are interested in conditions under which $S_t = \exp(X_t)$ is
conservative and a martingale. If such conditions are satisfied,
$\St$ may serve as the price process under the risk-neutral
measure in an arbitrage-free asset pricing model. The following
theorem gives sufficient and necessary conditions:

\begin{thm}\label{Thm:martingale}
Suppose $(X_t,V_t)$ satisfies Assumptions A1 and A2. Then the
following holds:
\begin{enumerate}[(a)]
\item \label{Thm:martingale_sub1}$\St$ is conservative if and only
if $F(0,0) = R(0,0) = 0$ and
\begin{equation}\label{Eq:conservative_integral}
\int_{0-}{\frac{d\eta}{R(0,\eta)}} = -\infty\;;
\end{equation}
\item \label{Thm:martingale_sub2}$\St$ is a martingale if and only
if it is conservative, $F(1,0) = R(1,0) = 0$ and
\begin{equation}\label{Eq:mg_integral}
\int_{0-}{\frac{d\eta}{R(1,\eta)}} = -\infty\;.
\end{equation}
\end{enumerate}
\end{thm}
\begin{rem}\label{Rem:Osgood}
The notation $\int_{0-}$ denotes an integral over an arbitrarily
small left neighborhood of $0$.
\end{rem}
By \eqref{Eq:FR_form} the condition $F(0,0) = R(0,0) = 0$ is
equivalent to $c = \gamma = 0$, i.e. obviously the killing rate
has to be zero for the process to be conservative. As will be seen
in the proof, the integral conditions
\eqref{Eq:conservative_integral} and \eqref{Eq:mg_integral} are
related to a uniqueness condition for non-Lipschitz ODEs, which
has been discovered by \citet{Osgood}.\\
The following Corollary gives easy-to-check sufficient conditions:
\begin{cor}\label{Cor:martingale}
Suppose $(X_t,V_t)$ satisfies Assumptions A1 and A2.
\begin{enumerate}[(a)]
\item If $F(0,0) = R(0,0) = 0$ and $\chi(0) < \infty$ then $\St$
is conservative. \item If $\St$ is conservative, $F(1,0) = R(1,0)
= 0$ and $\chi(1) < \infty$, then $\St$ is a martingale.
\end{enumerate}
\end{cor}

\begin{proof}
For a proof of \ref{Thm:martingale}\ref{Thm:martingale_sub1} we
refer to \cite [Th.~ 4.11]{Filipovic}. Statement
\ref{Thm:martingale}\ref{Thm:martingale_sub2}
can be shown in a similar way:\\
Since $(X_t,V_t)$ is Markovian, we have for all $0 \le s \le t$,
that
\[\EE[S_t|\cF_s] = S_s \exp\left(\phi(t-s,1,0) + V_s \psi(t-s,1,0)\right)\;.\]
We have assumed that $V_0 > 0$ a.s., such that $\St$ is a
martingale if and only if $\Xt$ is conservative and $\psi(t,1,0) =
\phi(t,1,0) \equiv 0$ for all $t \in \Rplus$.\\
We show Corollary~\ref{Cor:martingale} and the first implication
of \ref{Thm:martingale}\ref{Thm:martingale_sub2}: Suppose that
$\St$ is conservative and that $F(1,0) = R(1,0) = 0$. By
Theorem~\ref{Thm:Riccati} $\psi(t,1,w)$ solves the differential
equation
\begin{align}\label{Eq:Riccati_mg}
\pd{}{t}\psi(t,1,w) = R(1,\psi(t,1,w)), \qquad \psi(0,1,w) = w\;
\end{align}
for all $w \le 0$. Since $R(1,0) = 0$ it is clear that
$\widetilde{\psi}(t,1,0) \equiv 0$ satisfies this ODE for the
initial value $w = 0$. To deduce that $\widetilde{\psi}(t,1,0) =
\psi(t,1,0)$ however, we need to know whether the solution is
unique. Since $R(1,w)$ is continuously differentiable for $w < 0$,
it satisfies a Lipschitz condition on $(-\infty,0)$. If $\chi(1) <
\infty$, the Lipschitz condition can be extended to $(-\infty,0]$,
and $\psi(t,1,0) \equiv 0$ is the unique solution. Without the
assumption that $\chi(1) < \infty$, we substitute Lipschitz'
condition by Osgood's condition\footnote{See \citet{Osgood}}
\eqref{Eq:mg_integral}:
%\begin{equation}\label{Eq:Osgood}
%\int_0^{-\epsilon}\frac{d\eta}{R(1,\eta)} = \infty.
%\end{equation}
Suppose that \eqref{Eq:mg_integral} holds, and there exists a
non-zero solution $\widetilde{\psi}$ such that
$\widetilde{\psi}(t_1,1,0) < 0$ for some $t_1 > 0$. Then for all
$t < t_1$ such that $\psi$ remains non-zero on $[t,t_1]$ we have
(similarly to \eqref{Eq:eta_integral}) that
\begin{equation}\label{Eq:Osgood_interm}
\int_{\widetilde{\psi}(t_1,1,0)}^{\widetilde{\psi}(t,1,0)}
\frac{d\eta}{R(1,\eta)} = t - t_1\;.
\end{equation}
Assume that $t_0 \ge 0$ is the first point left of $t_1$ such that
$\widetilde{\psi}(t_0,1,w) = 0$. Letting $t \downarrow t_0$, the
left side of \eqref{Eq:Osgood_interm} tends to $-\infty$, whereas
the right side remains bounded, leading to a contradiction. We
conclude that $\psi(t,1,0) \equiv 0$ is the unique solution of
\eqref{Eq:Riccati_mg}. Finally equation
\eqref{Eq:Riccati_explicit} together with $F(1,0) = 0$ yields that
also $\phi(t,1,0) \equiv 0$ for all $t \in \Rplus$ and we
have shown that $\St$ is a martingale.\\

For the other direction of
\ref{Thm:martingale}\ref{Thm:martingale_sub2} note that $\St$
being a martingale implies that $\phi = \psi \equiv 0$ solve the
generalized Riccati equations and thus that $F(1,0) = R(1,0) = 0$.
It remains to show \eqref{Eq:mg_integral}. Assume that
\eqref{Eq:mg_integral} does not hold. Then, for each $t > 0$,
\eqref{Eq:Osgood_interm} with $t_1 = 0$ implicitly defines a
solution $\widetilde{\psi}(t,1,0)$ of the generalized Riccati
equation \eqref{Eq:Riccati_mg}, satisfying
$\widetilde{\psi}(t,1,0) < 0$ for all $t > 0$. By uniqueness of
the solution $\psi(t,1,w)$ for $w < 0$ and the flow property
\eqref{Eq:flow_prop}, we have $\widetilde{\psi}(t+s,1,0) =
\psi(t,1,\widetilde{\psi}(s,1,0))$ for $t,s$ small enough. Letting
$s \downarrow 0$ we obtain $\psi(t,1,0) = \widetilde{\psi}(t,1,0)
< 0$, which is a contradiction to $\psi \equiv 0$.
\end{proof}

We add now two assumptions to A1 and A2 and complete our
definition of an affine stochastic volatility model:

\begin{description}
\item[A3] The discounted price process $S_t = e^{X_t}$ is a
martingale. \item[A4] $R(u,0) \neq 0$ for some $u \in \RR$.
\end{description}

Assumption A4 excludes models where the distribution of $\Xt$ does
not depend at all on the volatility state $V_0$. In such a case we
can not speak of a true stochastic volatility model, and it will
be beneficial to avoid these degenerate cases. We are now ready to
give our definition of an affine stochastic volatility model:

\begin{defn}\label{Def:ASVM}
The process $(X_t,V_t)_{t \ge 0}$ is called an affine stochastic
volatility model, if it satisfies assumptions A1 -- A4.
\end{defn}

A simple consequence of this definition, that will often be used
is the following:

\begin{lem}\label{Lem:R_strictly_convex}
Let $(X_t,V_t)_{t \geq 0}$ be an affine stochastic volatility
model. Then $R(u,0)$ is a strictly convex function, satisfying
$R(0,0) = R(1,0) = 0$.
\end{lem}
\begin{proof}
From assumption A3 and Theorem~\ref{Thm:martingale} it follows
that $R(0,0) = R(1,0) = 0$. Lemma~\ref{Lem:psiphi_prop} implies
that $R(u,0)$ is either strictly convex or an affine function.
Assume it is affine. Then $R(u,0) = 0$ for all $u \in \RR$. This
contradicts A4, such that we conclude that $R(u,0)$ is a strictly
convex function.
\end{proof}

\section{Long-term asymptotics}\label{Sec:LongTerm}

In this section we study the behavior of an affine stochastic
volatility model as $t \to \infty$. We focus first on the
stochastic variance process $\Vt$. Under mild assumptions this
process will converge in law to its invariant distribution:

\subsection{Stationarity of the variance process}\label{Sec:stationary}

\begin{prop}\label{Prop:stationary}
Suppose that A1 and A2 hold, that $\chi(0) < 0$ and the L\'evy
measure $m$ satisfies the logarithmic moment condition
\[\int_{y > 1}{\left(\log y\right)\,m(dx,dy) < \infty}.\]
Then $\Vt$ converges in law to its unique invariant distribution
$L$, which has the cumulant generating function
\begin{equation}\label{Eq:cumulant_stationary}
l(w) = \int_w^0{\frac{F(0,\eta)}{R(0,\eta)}\,d\eta} \qquad (w \le
0)\;.
\end{equation}
\end{prop}
\citet{KellerSteiner} show that under the given conditions the
process $\Vt$ converges in law to a limit distribution $L$, whose
cumulant generating function can be represented by
\eqref{Eq:cumulant_stationary}. A short argument at the end of
this paragraph shows that the limit distribution is also the
unique invariant distribution of $\Vt$. First we make the
following definition: Given some affine stochastic volatility
model $(X_t,V_t)_{t \geq 0}$, we introduce the process
$(\widetilde{X}_t, \widetilde{V}_t)_{t \geq 0}$, defined as the
Markov process with the same transition probabilities as
$(X_t,V_t)_{t \geq 0}$, but started with $X_0 = 0$ and $V_0$
distributed according to $L$. We will refer to $(\widetilde{X}_t,
\widetilde{V}_t)_{t \geq 0}$ as the stochastic volatility model
$(X_t,V_t)_{t \geq 0}$ `in the stationary variance regime'. We
also define the associated price process $\widetilde{S}_t :=
\exp(rt + \widetilde{X}_t)$. As we discuss in
Section~\ref{Sec:Applications} the process
$(\widetilde{X}_t,\widetilde{V}_t)_{t \geq 0}$ can be related to
the pricing of forward-starting options, when the time until the
start
of the contract is large.\\
The cumulant generating function of
$(\widetilde{X}_t,\widetilde{V}_t)$ is given by
\begin{multline}\label{Eq:cgf_stationary}
\log \EE[e^{u\widetilde{X}_t + w \widetilde{V}_t}] = \log
\EE\left[\exp\left(\phi(t,u,w) + \widetilde{V}_0
\psi(t,u,w)\right)\right] = \phi(t,u,w) + l(\psi(t,u,w))\;.
\end{multline}
We verify now that $L$ is indeed an invariant distribution of
$\Vt$:
\begin{multline}
\EE\left[\exp\left(w \widetilde{V}_t\right)\right] =
\exp\left(\phi(t,0,w) + l(\psi(t,0,w))\right) = \\
= \exp\left(\int_0^t{F(0,\psi(s,0,w))}ds +
\int_{\psi(t,0,w)}^0{\frac{F(0,\eta)}{R(0,\eta)}\,d\eta}\right) = \\
=
\exp\left(\int_w^{\psi(t,0,w)}{\frac{F(0,\eta)}{R(0,\eta)}\,d\eta}
+ \int_{\psi(t,0,w)}^0{\frac{F(0,\eta)}{R(0,\eta)}\,d\eta}\right)
= \exp(l(w))\;,
\end{multline}
where we have used that under the conditions of the Proposition
above, $\psi(t,0,w)$ is a strictly monotone function converging to
$0$ as $t \to \infty$. (cf. \citet{KellerSteiner}). To see that
$L$ is unique, assume that there exists another invariant
distribution $L'$, and let $(V'_t)_{t \geq 0}$ be the variance
process started with $V'_0$ distributed according to $L'$. Again
we use that $\phi(t,u,w) \to l(w)$ and $\psi(t,0,w) \to 0$ as $t
\to \infty$ (see \citet{KellerSteiner}), and get that
\begin{multline*}
\lim_{t \to \infty} \EE\left[\exp(wV'_t)\right] = \EE\left[\lim_{t
\to \infty} \exp\left(\phi(t,0,w) + V'_0 \psi(t,0,w)\right)\right]
= \EE[\exp(l(w))] = e^{l(w)}\;,
\end{multline*}
for all $w \le 0$, in contradiction to the invariance of $L'$.

\subsection{Long-term behavior of the log-price process}

We have seen that $\Vt$ converges to a limit distribution, but we
do not expect the same for the log-price process $\Xt$.
Nevertheless, it can be shown that the rescaled cumulant
generating function $\frac{1}{t} \log \EE\left[e^{X_t u}\right]$
converges under suitable conditions to a limit $h(u)$, that is
again the cumulant generating function of some infinitely
divisible random variable. This result can be interpreted such,
that for large $t$ the marginal distributions of $\Xt$ are `close'
to the marginal distributions of a L\'evy process with
characteristic exponent $h(u)$. Furthermore, $h(u)$ can be
directly obtained from the functions $F$ and $R$, without
knowledge of the explicit forms of $\phi$ and $\psi$. We start
with a preparatory Lemma:

\begin{lem}\label{Lem:existence}
Let $(X_t,V_t)_{t \geq 0}$ be an affine stochastic volatility
model and suppose that $\chi(0) < 0$ and $\chi(1) < 0$. Then there
exist a maximal interval $I$ and a unique function $w \in C(I)
\cap C^1(I^\circ)$, such that
\[R(u,w(u)) = 0 \qquad \text{for all} \quad u \in I\]
and $w(0) = w(1) = 0$.\\
Moreover it holds that $[0,1] \subseteq I$, $w(u) < 0$ for all $u
\in (0,1)$; $w(u)
> 0$ for all $u \in I \setminus [0,1]$; and
\begin{equation}\label{Eq:partialR_negative}
\pd{R}{w}(u,w(u)) < 0
\end{equation}
for all $u \in I^\circ$.
\end{lem}

We show Lemma~\ref{Lem:existence} together with the next result,
which makes the connection to the qualitative properties of the
generalized Riccati equations.

\begin{lem}\label{Lem:aux}
\begin{enumerate}[(a)]
\item For each $u \in I^\circ$, $w(u)$ is an asymptotically stable
equilibrium point of the generalized Riccati equation
\eqref{Eq:gen_Riccati_sub2}. \item For $u \in I^\circ$, there
exists at most one other equilibrium point $\widetilde{w}(u) \neq
w(u)$, and if it exists, it is necessarily unstable and satisfies
$\widetilde{w}(u)
> \max(0,w(u))$. \item For $u \in \RR \setminus I$, no
equilibrium point exists.
\end{enumerate}
\end{lem}

\begin{proof}
Define $L = \set{(u,w) : R(u,w) \le 0}$. As the level set of the
closed convex function $R$, it is a closed and convex set. For all
$u \in \RR$, define $w(u) = \inf \set{w: (u,w) \in L}$, and $I =
\set{u \in \RR: w(u) < \infty}$. Clearly $w(u)$ is a continuous
convex function, and $I$ a subinterval of $\RR$. We will now show
that $w(u)$ and $I$ satisfy all properties stated in
Lemma~\ref{Lem:existence}. By assumption A3 and
Theorem~\ref{Thm:martingale}, $R(0,0) = R(1,0) = 0$; together with
Lemma~\ref{Lem:psiphi_prop} it follows that the set $[0,1] \times
(-\infty,0]$ is contained in $\dom\,R$. Since $R(u,0)$ is by
Lemma~\ref{Lem:R_strictly_convex} strictly convex, and also
$\chi(u)$ is convex, we deduce that $R(u,0) < 0$ and
$\pd{R}{w}(u,0) = \chi(u) < 0$ for all $u \in (0,1)$. In addition
$R(u,w)$, as a function of $w$, is either affine or strictly
convex, such that there exists a unique point $w(u)$, where
$R(u,w(u)) = 0$, and necessarily $\pd{R}{w}(u,w(u)) < 0$. It is
clear that for $u \in (0,1)$ $w(u)$ coincides with the function
defined above, and that $w(u) < 0$. At $u = 0$ we have that
$R(0,0) = 0$ and $\chi(0) < 0$, implying that $w(0) = 0$. A
symmetrical
argument at $u = 1$ shows that $w(1) = 0$, and thus that $[0,1] \subseteq I$.\\
We show next that $w(u) \in C^1(I^\circ)$: Define $u_+ = \sup I$,
and $w_+ = \lim_{u \uparrow u_+}w(u)$; $u_-, w_-$ are defined
symmetrically at the left boundary of $I$. Note that $u_\pm$ and
$w_\pm$ can take infinite values. Define the open set
\[K := \set{\left(\lambda u_- + (1 - \lambda) u_+, w\right)\,:\,\lambda \in (0,1),\,w < \lambda w_- + (1 - \lambda)w_+}\;.\]
Lemma~\ref{Lem:psiphi_prop} implies that $K$ is contained in the
interior of $\dom\,R$. On the other hand, the graph of $w$,
restricted to $I^\circ$, i.e. the set $\set{(u,w(u)): u \in
I^\circ}$, is clearly contained in $K$. Since $R$ is by
Lemma~\ref{Lem:psiphi_prop} an analytic function in the interior
of its effective domain, the implicit function theorem implies
that $w(u) \in C^1(I^\circ)$. In addition it follows that
$\pd{R}{w}(u,w(u)) \neq 0$ for all $u \in I^\circ$, such that the
assertion $\pd{R}{w}(u,w(u)) < 0$, which we have shown above for
$u \in (0,1)$, can be extended to all of $I^\circ$. The claim that
$w(u) > 0$ for $u \in I \setminus [0,1]$ can easily be derived
from the convexity of $w(u)$, and the fact that $w(u) < 0$
inside $(0,1)$ and $w(0) = w(1) = 0$.\\

We have now proved most part of Lemma~\ref{Lem:existence} (except
for the uniqueness), and turn towards Lemma~\ref{Lem:aux}: Since
$R(u,w(u)) = 0$ and $\pd{R}{w}(u,w(u)) < 0$ for all $u \in
I^\circ$, $w(u)$ must be an asymptotically stable equilibrium
point of the generalized Riccati
equation~\ref{Eq:gen_Riccati_sub2}, showing \ref{Lem:aux}a. Assume
now that for some $u \in I^\circ$ there exists a point
$\widetilde{w}(u) \neq w(u)$ such that $R(u,\widetilde{w}(u)) =
0$. By Lemma~\ref{Lem:psiphi_prop}, $R(u,w)$ is, as a function of
$w$, either strictly convex or affine. If it is affine, it has a
unique root, and $\widetilde{w}(u)$ cannot exist. If it is
strictly convex, there can exist a single point $\widetilde{w}(u)$
other than $w(u)$, such that $R(u,\widetilde{w}(u)) = 0$.
Necessarily $\widetilde{w}(u) > w(u)$ and
$\pd{R}{w}(u,\widetilde{w}(u)) > 0$. This shows that
$\widetilde{w}(u)$ is an unstable equilibrium point of the
generalized Riccati equation for $\psi$. In addition
$\widetilde{w}(u) > w(u)$, and in particular the fact that
$\widetilde{w}(0)
> 0$ and $\widetilde{w}(1)
> 0$ shows the uniqueness of $w(u)$ in the sense of
Lemma~\ref{Lem:existence}. To see that $\widetilde{w}(u) >
\max(0,w(u))$, note that we only have to show that
$\widetilde{w}(u)
> 0$, whenever $w(u) < 0$. This is the case only for $u \in
(0,1)$. Assume that $\widetilde{w}(u) \le 0$ for $u \in (0,1)$.
Then the convexity of $R$ and $\pd{R}{w}(u,\widetilde{w}(u)) > 0$
would imply that $R(u,0) \ge 0$ for some $u \in (0,1)$. This is
impossible by Lemma~\ref{Lem:R_strictly_convex}, and we have shown
\ref{Lem:aux}b. Finally \ref{Lem:aux}c follows directly from the
definition of $w(u)$ as $w(u) = \inf \set{w: (u,w) \in L}$ and $I$
as the effective domain of $w(u)$.
\end{proof}

We are now ready to show our main result on the long-term
properties of the log-price process $\Xt$.

\begin{thm}\label{Thm:wm_convergence}
Let $(X_t,V_t)_{t \geq 0}$ be an affine stochastic volatility
model and suppose that $\chi(0) < 0$ and $\chi(1) < 0$. Let $w(u)$
be given by Lemma~3.2 and define
\[h(u) = F(u,w(u)), \qquad J = \set{u \in I: F(u,w(u)) < \infty}\;.\]
Then $[0,1] \subseteq J \subseteq I$; $w(u)$ and $h(u)$ are
cumulant generating functions of infinitely divisible random
variables and
\begin{subequations}\label{Eq:limit_psiphi}
\begin{align}
\lim_{t \to \infty} \psi(t,u,0) &= w(u) \quad \text{for all} \quad u \in I\;;\label{Eq:limit_psiphi_sub1}\\
\lim_{t \to \infty} \frac{1}{t}\phi(t,u,0) &= h(u) \quad \text{for
all} \quad u \in J\;.\label{Eq:limit_psiphi_sub2}
\end{align}
\end{subequations}
\end{thm}

\begin{cor}\label{Cor:phipsi_estimate}
Under the conditions of Theorem~\ref{Thm:wm_convergence}, the
following holds:
\begin{subequations}\label{Eq:estimate_psiphi}
\begin{align}
\sup_{u \in [0,1]}\left|\psi(t,u,0) - w(u)\right| &\le C \exp(-\mathfrak{X} \cdot T)\;; \label{Eq:phipsi_est_sub1}\\
\sup_{u \in [0,1]}\left|\frac{1}{t}\phi(t,u,0) - h(u)\right| &\le
\Omega C \exp(-\mathfrak{X} \cdot T)\;; \label{Eq:phipsi_est_sub2}
\end{align}
\end{subequations} for some constant $C$, and with
\[\mathfrak{X} = \inf_{u \in [0,1]}|\chi(u)|\quad \text{and} \quad \Omega = \sup_{u \in [0,1]}\left.\pd{}{w}F(u,w)\right|_{w = 0}\,\]
\end{cor}

\begin{proof}
Let $u \in [0,1]$. By Lemma~\ref{Lem:existence} $(u,w(u)) \in
[0,1] \times (-\infty,0]$. By Theorem~\ref{Thm:martingale} $F(0,0)
= F(1,0) = 0$, such that Lemma~\ref{Lem:psiphi_prop} guarantees
that $[0,1] \times (-\infty,0] \subseteq \dom\,F$. It follows that
$[0,1] \subseteq J$. Define
\[z(t,u) = \psi(t,u,0) - w(u)\;.\]
Inserting into the generalized Riccati equation
\ref{Eq:gen_Riccati_sub2},
\begin{equation*}
\pd{}{t} z(t,u) = R(u,\psi(t,u,0)) = R(u,\psi(t,u,0)) - R(u,w(u)),
\quad \text{and} \quad z(0,u) = w(u)\;.
\end{equation*}
If $\psi(t,u,0) \le 0$ we can bound the right hand side by
\[R(u,\psi(t,u,0)) - R(u,w(u)) \le z(t,u) \pd{R}{w}(u,0) = z(t,u) \chi(u),\]
using convexity of $R$. By Gronwall's inequality
\[ z(t,u) \le |w(u)| \exp\left(\chi(u) t\right)\;.\]
Since $\chi$ is convex, $\chi(0) < 0$ and $\chi(1) < 0$, we have
shown \eqref{Eq:phipsi_est_sub1}. The estimate
\begin{multline*}
\left|\phi(t,u) - h(u)\right| = \\
 = \left|\frac{1}{t}\int_0^t{\left(F(u,\phi(s,u)) -
F(u,w(u))\right)\;ds}\right|  \le \left|\pd{F}{w}(u,0)\right|
\cdot |\psi(t,u) - w(u)|
\end{multline*}
yields \eqref{Eq:phipsi_est_sub2} and we have shown
Corollary~\ref{Cor:phipsi_estimate}.\\
Let now $u \in I^\circ \setminus [0,1]$. Combining
Lemma~\ref{Lem:R_strictly_convex} and Lemma~\ref{Lem:aux} we have
that $R(u,w) > 0$ for all $w \in [0,w(u))$,  and $R(u,w(u)) = 0$.
It follows that the initial value $\psi(0,u,0)=0$ is in the basin
of attraction of the stable equilibrium point $w(u)$ and thus that
$\psi(t,u,0)$ is strictly increasing and converging to $w(u)$. An
additional argument may be needed at the boundary of $I$: Let $u_+
= \sup I$ and assume that $u_+ \in I$ (i.e. $I$ is right-closed).
Since $(u_+,w) \in \dom\,R$ for all $w \le w(u_+)$, we can define
$\pd{R}{w}(u_+,w(u_+))$ at least as a limit for $w \uparrow
w(u_+)$. By Lemma~\ref{Lem:existence} either
$\pd{R}{w}(u_+,w(u_+)) < 0$ or $\pd{R}{w}(u_+,w(u_+)) = 0$. In the
first case we can argue as in the interior of $I$ that $w(u_+)$ is
an asymptotically stable equilibrium point. In the second case we
use once more that by Lemma~\ref{Lem:psiphi_prop} $R(u_+,w)$ is,
as a function of $w$, either strictly convex or affine. If it is
affine, it must be equal to $0$, and thus $R(u_+,0) = 0$, in
contradiction to Lemma~\ref{Lem:R_strictly_convex}. Hence it is
strictly convex, and attains its minimum at $w(u_+)$. This implies
that $R(u_+,w) > 0$ for all $w \in [0,w(u_+))$ and we conclude
that $\psi(t,u_+,0)$ converges to $w(u_+)$.\footnote{Even though
$\psi(t,u_+,0)$ converges to $w(u_+)$, note that $w(u_+)$ is not a
stable equilibrium point in the usual sense. This is due to the
fact that solutions from a \emph{right-}neighborhood $N \cap
(w(u_+),\infty)$ will \emph{diverge} from $w(u_+)$ to $+\infty$.}
For $u_- =
\inf I$, a symmetrical argument applies.\\
Assertion \eqref{Eq:limit_psiphi_sub2} follows immediately from
the representation \eqref{Eq:Riccati_explicit}, and
\[\lim_{t \to \infty}\frac{1}{t}\phi(t,u,0) = \lim_{t \to \infty}\frac{1}{t}\int_0^t{F(u,\psi(s,u,0))\;ds} = F(u,w(u))\]
for all $u \in J$.\\
We have shown that the sequence of infinitely divisible cumulant
generating functions $\psi(t,u,0)$ converges on $I$ to a function
$w(u)$ that is continuous in a right neighborhood of $0$. This is
sufficient to imply that $w(u)$ is again the cumulant generating
function of an infinitely divisible random variable (See
\citet[VIII.1, Example~(e)]{Feller} for the convergence part, and
\citet[Lemma~7.8]{Sato} for the infinite divisibility.). The same
argument can be applied to $\phi$ and $h(u)$, and we have shown
Theorem~\ref{Thm:wm_convergence}.
\end{proof}

\section{Moment explosions}\label{Sec:MomentExplosion}
In this section we continue to study the time evolution of moments
$\EE[S_t^u] = \EE[e^{X_t u}]$ of the price process in an affine
stochastic volatility model. We are interested in the phenomenon
that in a stochastic volatility model, moments of the price
process can explode (become infinite) in finite time. For
stochastic volatility models of the CEV-type -- a class including
the Heston model, but no models with jumps -- moment explosions
have been studied by \citet{Piterbarg} and \citet{LionsMusiela}.
In the context of option pricing, an interesting result of
\citet{Lee} connects the existence of moments of the stock price
process to the steepness of the smile for deep in-the-money or
out-of-the-money options. Our first result shows that in an affine
stochastic volatility model a simple explicit expression for the
time of moment explosion can be given:

\subsection{Moment explosions}
By definition, the $u$-th moment of $S_t$, i.e. $\EE[S_t^u]$ is
given by $S_0^u \exp\left(\phi(t,u,0) + V_0 \psi(t,u,0)\right)$.
We define the \textbf{time of moment explosion} for the moment of
order $u$ by
\[T_*(u) = \sup \set{t : \EE[S_t^u] < \infty}\;.\]
It is obvious from the Markov property that $\EE[S_t^u]$ is finite
for all $t < T_*(u)$ and infinite for all $t > T_*(u)$. As in the
previous section, the main result follows from a qualitative
analysis of the generalized Riccati equations
\eqref{Eq:gen_Riccati}.

\begin{thm}\label{Thm:moment_explosions}
Suppose the conditions of Theorem~\ref{Thm:wm_convergence} hold.
Define $J = \set{u \in I: F(u,w(u)) < \infty}$,
\begin{align*}
f_+(u) &:= \sup \set{w \ge 0 : F(u,w) < \infty}\;,\\
r_+(u) &:= \sup \set{w \ge 0 : R(u,w) < \infty}\;,
\end{align*}
and suppose that $F(u,0) < \infty$, $R(u,0) < \infty$ and $\chi(u)
< \infty$.
\begin{enumerate}[(a)]
\item If $u \in J$, then
\[T_*(u) = +\infty\;.\] \label{Eq:moments_sub1}
\item  If $u \in \RR \setminus J$, then
\begin{equation*}\label{Eq:explosion_integral}
T_*(u) = \int_0^{\min(f_+(u),r_+(u))} {\frac{d\eta}{R(u,\eta)}}\;.
\end{equation*}\label{Eq:moments_sub2}
\end{enumerate}
If $F(u,0) = \infty$, $R(u,0) = \infty$ or $\chi(u) = \infty$ then
\begin{enumerate}[(c)]
\item  \[T_*(u) = 0\;.\] \label{Eq:moments_sub3}
\end{enumerate}
\end{thm}

%The next Corollary will be used later as an auxiliary result:
%
%\begin{cor}\label{Cor:increasing_moment}
%Let $u \in \RR \setminus [0,1]$. Then $\EE[S_t^u]$ is an
%increasing function of $T$.
%\end{cor}
%
\begin{proof}
Suppose that $u \in J$. Then Theorem~\ref{Thm:wm_convergence}
implies that both $\psi(t,u,0)$ and $\phi(t,u,0)$ are finite for
all $t \ge 0$. This proves \eqref{Eq:moments_sub1}. Let now $u \in
\RR \setminus J$, $F(u,0) < \infty$, $R(u,0) < \infty$ and
$\chi(u) < \infty$. To prove \eqref{Eq:moments_sub2} we start by
analyzing the maximal lifetime of solutions to the generalized
Riccati equation
\begin{equation}\label{Eq:gen_Riccati_repeat}
\pd{}{t}\psi(t,u,0) = R(u,\psi(t,u,0)), \quad \psi(0,u,0) = 0\;.
\end{equation}
Define $M = [0,r_+(u))$ and note that $R(u,.) \in C(M)$. Since $u
\not \in [0,1]$, Lemma~\ref{Lem:R_strictly_convex} implies that
$R(u,0)
> 0$. It is clear, that at least a local solution $\psi(t,u,0)$ to
the ODE exists, which satisfies $0 \le \psi(t,u,0) \le r_+(u)$ and
is an increasing function of $t$ as long as it can be continued.
Using a standard extension theorem (e.g.
~\citet[Lem.~I.3.1]{Hartman}) the local solution $\psi(t,u,0)$ has
a maximal extension to an interval $[0,T(u))$, such that one of
the following holds:
\begin{enumerate}[(i)]
\item \label{Eq:extension_case1}$T(u) = \infty$, or \item
\label{Eq:extension_case2} $T(u) < \infty$ and $\psi(t,u,0)$ comes
arbitrarily close to the boundary of $M$, i.e.
\[\limsup_{t \to T(u)} \psi(t,u,0) = r_+(u)\;.\]
\end{enumerate}
Consider case \eqref{Eq:extension_case1}. Since $\psi$ is
increasing, its limit for $t \to \infty$ exists, but can be
infinite. Suppose $\lim_{t \to \infty} \psi(t) = \alpha < \infty$.
Then $\alpha$ must be a stationary point, i.e. $R(u,\alpha) = 0$,
but this is impossible by Lemma~\ref{Lem:aux}. The case that
$\alpha = \infty$ is only possible if $r_+(u) = \infty$, such that
in this case $\lim_{t \to T(u)}\psi(t,u,0) = r_+(u)$. Consider
case \eqref{Eq:extension_case2}. Since $\psi$ is increasing the
limes superior can be replaced by an ordinary limit and we get
$\lim_{t \to T(u)}\psi(t,u,0) = r_+(u)$ as before.\\

Let now $T_n$ be a sequence such that $T_n \uparrow T(u)$. By
\eqref{Eq:eta_integral} it holds that
\begin{equation}\label{Eq:eta_integral_2}
    \int_0^{\psi(T_n,u,0)}{\frac{d\eta}{R(u,\eta)}ds} = T_n\;.
\end{equation}
Letting $n \to \infty$ we obtain that $T(u) =
\int_0^{r_+(u)}{\frac{d\eta}{R(u,\eta)}ds}$.\\

We can write the time of moment explosion $T_*(u)$ as the maximum
joint lifetime of $\phi(t,u,0)$ and $\psi(t,u,0)$, i.e. $T_*(u) =
\sup \set{t \ge 0: \phi(t,u,0) < \infty \wedge \psi(t,u,0) <
\infty}$. By the integral representation
\eqref{Eq:Riccati_explicit} it is clear that if $f_+(u) \ge
r_+(u)$, $\phi(t,u,0)$ is finite whenever $\psi(t,u,0)$ is finite
and $T_*(u) = T(u)$. If $f_+(u) < r_+(u)$ then $\psi(T_*(u),u,0) =
f_+(u)$. Inserting into the
representation \eqref{Eq:eta_integral_2} yields \eqref{Eq:moments_sub2}.\\
For assertion (c), let $F(u,0) = \infty$, $R(u,0) = \infty$, or
$\chi(u) = \infty$. In the first case, $\phi(t,u,0)$ does not
exist beyond $t = 0$. In the other cases no local solution to the
generalized Riccati equation \eqref{Eq:gen_Riccati_repeat} exists,
such that $\psi(t,u,0)$ explodes immediately.
\end{proof}

\subsection{Moment explosions in the stationary variance regime}
In Section~\ref{Sec:stationary} we have introduced
 $(\widetilde{X}_t, \widetilde{V}_t)_{t \geq 0}$ as the model in
the stationary variance regime. The moment explosions of this
process can be analyzed in a similar manner as above. We define
the time of moment explosion in the stationary variance regime by
\[T^S_*(u) := \sup \set{T \geq 0 : \EE[\widetilde{S}_T^u] <
\infty}\;;\] the superscript `S' stands for `stationary'.

%Analogously to Section~\ref{Sec:smile}, we also define the
%critical moment functions under stationary variance, by
%\begin{align*}
%u_+^S(T) &:= \sup \set{u \ge 1 :
%\EE[\widetilde{S}_T^u] <
%\infty}\;,\\
%u_-^S(T) &:= \inf \set{u \le 0 :
%\EE[\widetilde{S}_T^u]}\;.
%\end{align*}

The analogue to Theorem~\ref{Thm:moment_explosions} is the
following result:

\begin{thm}\label{Thm:moment_explosions_stationary}
Suppose the conditions of Theorem~\ref{Thm:wm_convergence} hold.
Define $f_+(u),r_+(u)$ as in Theorem~\ref{Thm:moment_explosions},
and in addition
\[l_+ := \sup \set{w > 0 : l(w) < \infty}\;.\]
Suppose that $F(u,0) < \infty$, $R(u,0) < \infty$ and $\chi(0) <
\infty$.
\begin{enumerate}[(a)]
\item If $u \in J$ and $w(u) \le l_+$, then
\[T_*^S(u) = +\infty\;.\]
\item If $u \in \RR \setminus J$ or $w(u) > l_+$, then
\[T_*^S(u) = \int_0^{\min(f_+(u),r_+(u),l_+)} {\frac{d\eta}{R(u,\eta)}}\;.\]
\end{enumerate}
If $F(u,0) = \infty$, $R(u,0) = \infty$ or $\chi(0) = \infty$,
then
\begin{enumerate}[(c)]
\item \[T_*^S(u) = 0\;.\]
\end{enumerate}
\end{thm}

\begin{cor}\label{Cor:difference_stationary}
Under the conditions of
Theorem~\ref{Thm:moment_explosions_stationary},
\[T_*^S(u) \le T_*(u), \qquad \text{for all $u \in \RR$}\]
\end{cor}

\begin{proof}
By equation~\eqref{Eq:cgf_stationary}, the moment
$\EE[\widetilde{S}_t^u]$ is given by
\[\EE[\widetilde{S}_t^u] = \exp\left(\phi(t,u,0) + l(\psi(t,u,0))\right)\;.\]
This expression is finite, if $\phi(t,u,0)$ and $\psi(t,u,0)$ are
finite, \emph{and} if $\psi(t,u,0) < l_+$. It is infinite if
$\phi(t,u,0)$ or $\psi(t,u,0)$ are infinite, or if $\psi(t,u,0)
> l_+$. The rest of the proof can be carried out as for
Theorem~\ref{Thm:moment_explosions}. Note, that now even for $u
\in J$, the moment can explode, if $l_+$ is reached by
$\psi(t,u,0)$ before the stationary point $w(u)$.
Corollary~\ref{Cor:difference_stationary} follows easily by
comparing the range of integration and the conditions for case (a)
and (b) between Theorem~\ref{Thm:moment_explosions} and
Theorem~\ref{Thm:moment_explosions_stationary}.
\end{proof}

\section{Applications}\label{Sec:Applications}

%\subsection{Saddlepoints}
%Based on the long-term behaviour of $\EE^[(S_t)/(S_0)^u]$,
%\cite{Lewis} derives approximations for the implied volatility of
%several diffusion models for large time-to maturity. The
%
%Write the price of a European call as Fourier Integral, and use a
%saddlepoint approximation:
%\begin{align*}
%\frac{1}{S_0}C(T,\xi) &= 1 - \frac{e^{(1 - u_*) \xi}}{2 \pi} \int_{-\infty}^\infty{\frac{e^{-i z \xi} \exp\left(\Phi_T(u_* + iz)\right)}{(z + i(1 - u_*))(z - iu_*)}\,dz} = \\
%&= 1 - \frac{\exp\left((1 - u_*)\xi + T m(u_*) + C\right)}{2
%u_*(1-u_*) \pi \sqrt{T}}\sqrt{\frac{1}{2 \pi m''(u_*)}} + \\ & +
%\mathcal{O}\left(\frac{1}{T}\right)
%\end{align*}

\subsection{Smile behavior at extreme strikes}

In the preceding section, we have kept $u$ fixed, and looked at
the first time $T_*(u)$ that the moment $\EE[S_t^u]$ becomes
infinite. It will now be more convenient to reverse the roles of
$T$ and $u$, and for a given time $t$ to define the \textbf{upper
critical moment} by
\[u_+(t) = \sup \set{u \ge 1 : \EE[S_t^u] < \infty} = \sup \set{u \ge 1 : T_*(u) < t}\;,\]
and the \textbf{lower critical moment} by
\[u_-(t) = \inf \set{u \le 0 : \EE[S_t^u] < \infty} = \inf \set{u \le 0 : T_*(u) < t} \;.\]

It is seen that $u_-(T)$ and $u_+(T)$ can be defined as the
generalized inverse of $T_*(u)$ on $(-\infty,0]$ and $[1,\infty)$
respectively. In addition it is easily derived from Jensen's
inequality, that
\begin{alignat*}{3}
\EE[S_t^u] &<& \infty \qquad &\text{for all}\; u \in (u_-(t),u_+(t)), \quad \text{and}\\
\EE[S_t^u] &=& \infty \qquad &\text{for all}\; u \in \RR \setminus
[u_-(t),u_+(t)]\;.
\end{alignat*}

The results of \citet{Lee} relate the explosion of moments to the
'wing behavior' of the implied volatility smile, i.e. the shape of
the smile for strikes that are deep in-the-money or
out-of-the-money. To give a precise statement, let $\xi$ be the
log-moneyness, which for a European option with time-to-maturity
$T$ and strike $K$ is given by $\xi = \log\left(\frac{K}{e^{r T}
S_0}\right)$.

\begin{prop}[Lee's moment formula]\label{Prop:Lee}
Let $V(T,\xi)$ be the implied Black-Scholes-Variance of a European
call with time-to-maturity $T$ and log-moneyness $\xi$. Then
\[\limsup_{\xi \to -\infty}\frac{V(T,\xi)}{|\xi|} = \frac{\varsigma(-u_-(T))}{T}\]
and
\[\limsup_{\xi \to \infty}\frac{V(T,\xi)}{|\xi|} = \frac{\varsigma(u_+(T) - 1)}{T}\]
where $\varsigma(x) = 2 - 4 \left(\sqrt{x^2 + x} - x\right)$ and
$u_\pm(T)$ are the critical moment functions.
\end{prop}

The function $\varsigma$ is strictly decreasing on $\Rplus$,
mapping $0$ to $2$, and $\infty$ to $0$. Thus for fixed
time-to-maturity $T$, the steepness of the smile is decreasing
with $|u_\pm(T)|$. A finite critical moment $u_\pm(T)$ implies
asymptotically linear behavior of $V(T,\xi)$ in $\xi$, and an
infinite critical moment implies sublinear behavior of $V(T,\xi)$.
It is also evident that $u_-(T)$ determines the 'left' side of the
volatility smile, also known as small-strike, in-the-money-call or
out-of-the-money-put side; $u_+(T)$ determines the 'right' side,
or large-strike, out-of-the-money-call, in-the-money-put side.
Finally we mention that Lee's result has been extended and
strengthened by \citet{Friz} from a `$\limsup$' to a genuine limit
under conditions related to regular variation of the underlying
distribution function.

\subsection{Forward-smile behavior}

The forward smile is derived from the prices of forward-start
options. For a forward-start call option -- all options we
consider are European -- a start date $\tau$, a strike date $T +
\tau$ and a moneyness ratio $M$ are agreed upon today (at time $t
= 0$). The option then yields at time $T + \tau$ a payoff of
$\left(\frac{S_{T + \tau}}{S_\tau} -  M\right)_+$, i.e. the
relative return over the time period from $\tau$ to $\tau + T$,
reduced by $M$ and floored at $0$. Under the pricing measure the
value of such an option at $t = 0$ is given by
\begin{equation}\label{Eq:fw_start_price}
e^{-r (T + \tau)} \EE\left[\left(\frac{S_{T + \tau}}{S_\tau} -
M\right)_+\right] = e^{-\tau r}\EE\left[\left(e^{X_{T + \tau} -
X_\tau} - e^\xi\right)_+\right]\;,
\end{equation}
where we define the log-moneyness $\xi$ of a forward-start option
as $\xi = \log M + rT$. Forward-start options are not just
interesting in their own right, but are used as building blocks of
more complex derivatives, such as Cliquet options (see
\citet[Chapter~10]{Gatheral}).\\
% As evident by
%\eqref{Eq:fw_start_price}, the value of a forward-start option
%will depend on the transitional distribution of the price process
%$\St$ between the time-points $\tau$ and $\tau + T$, and not on
%the marginal distributions, like the value of plain vanilla
%options. Thus even models that are perfectly calibrated to today's
%implied volatility smile can yield vastly different prices for
%forward starting options.\\
Analogously to plain vanilla options, we can define the
\textbf{implied forward volatility} $\sigma(\tau,T,\xi)$, by
comparing the forward option price to the price of an option with
identical payoff in the Black-Scholes model. Note that the implied
forward volatility depends also on $\tau$, the starting time of
the contract. For $\tau = 0$, the implied volatility of a plain
vanilla option is retrieved. More interesting is the behavior for
$\tau > 0$. Intuitively, we expect the implied volatility (and the
option price) to increase with $\tau$ in a stochastic volatility
model, since the uncertainty of the variance $V_\tau$ at the
starting date of the option has to be priced in. In an affine
stochastic volatility model, it will be seen that under mild
conditions, the implied forward volatilities $\sigma(\tau,T,\xi)$
actually converge to a limit as $\tau \to \infty$. Not
surprisingly, this behavior is related to the convergence of $\Vt$
to its invariant distribution. In the limit $\tau \to \infty$, the
pricing of a forward-start option is equivalent to the pricing of
a plain vanilla option \emph{in the stationary variance regime}
(cf. Section~\ref{Sec:stationary}).

\begin{prop}\label{Prop:limit_smile}
Let $(X_t,V_t)_{t \geq 0}$ be an affine stochastic volatility
model, satisfying the conditions of
Proposition~\ref{Prop:stationary}. Let $\sigma(\tau,T,\xi)$ be the
implied forward volatility in this model. Then
\[\lim_{\tau \to \infty} \sigma(\tau,T,\xi) = \widetilde{\sigma}(T,\xi)\;,\]
where $\widetilde{\sigma}(T,\xi)$ is the implied volatility of a
European call with payoff $\left(e^{\widetilde{X}_T} -
e^\xi\right)_+$, and $\widetilde{X}_T$ is the log-price process of
the model in the stationary variance regime.
\end{prop}

\begin{proof}We can write the price
of a forward-start call as
\begin{equation*}\label{Eq:rewrite_fw_price}
C(\tau,T,\xi) = e^{-\tau r}\EE\left[\left(e^{X_{T + \tau} -
X_\tau} - e^\xi\right)_+\right] = e^{-r\tau}
\EE\left[\EE^{(0,V_\tau)}\left[\left(e^{X_T} -
e^\xi\right)_+\right]\right]\;.
\end{equation*}
Denote by $C^\text{BS}(T,\xi,\sigma)$ the (plain vanilla) call
price in a Black-Scholes model with volatility $\sigma$ and the
normalization $S_0 = 1$. It is easy to see that the price of a
forward-start option in the Black-Scholes model is just the
discounted plain vanilla price, i.e.
$C^\text{BS}(\tau,T,\xi,\sigma) = e^{-r
\tau}C^\text{BS}(T,\xi,\sigma)$. By definition, the implied
forward volatility of the call $C(\tau,T,\xi)$ satisfies
\[C^\text{BS}(T,\xi,\sigma(\tau,T,\xi)) =  e^{r \tau}C(\tau,T,\xi)  = \EE\left[\EE^{(0,V_\tau)}\left[\left(e^{X_T} -
e^\xi\right)_+\right]\right]\;.\] Taking the limit $\tau \to
\infty$ on both sides we obtain
\[C^\text{BS}(T,\xi,\lim_{\tau \to \infty}\sigma(\tau,T,\xi)) = \EE\left[\lim_{\tau \to \infty} \EE^{(0,V_\tau)}\left[\left(e^{X_T} -
e^\xi\right)_+\right]\right] = \EE\left[\left(e^{\widetilde{X}_T}
- e^\xi\right)_+\right]\;,\] using dominated convergence. It is
well known that the above equation allows a unique solution in
terms of the Black-Scholes implied volatility, and we get
$\widetilde{\sigma}(T,\xi) = \lim_{\tau \to
\infty}\sigma(\tau,T,\xi)$.
\end{proof}

Combining Lee's moment formula with our results on moment
explosions under the stationary variance regime
(Theorem~\ref{Thm:moment_explosions_stationary}), asymptotics of
$\widetilde{\sigma}(T,\xi)$ for $\xi \to \pm \infty$ can be
derived.

\section{Examples}\label{Sec:Examples}

\subsection{The Heston model with and without jumps}\label{Sec:Heston}

\begin{figure}[tbp]
  \includegraphics[height=.45\textheight]{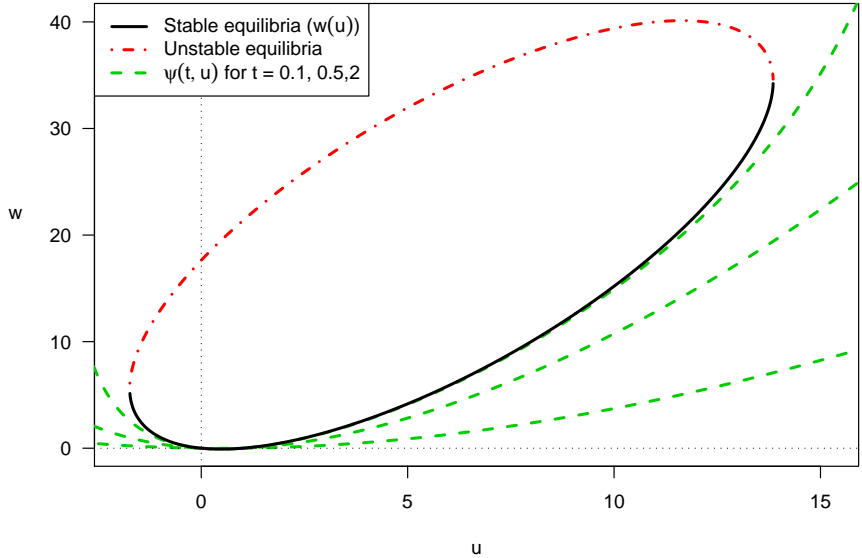}\\[-5pt]
  \caption{{\Small This plot shows the stable and unstable equilibria of the generalized Riccati equation of a Heston model with parameters
  $\rho=-0.7165$, $\zeta = 0.3877$, $\lambda = 1.3253$ and $\theta
  = 0.0354$
  (taken from \citet[Table~3.2]{Gatheral}). It can be seen how the solutions $\psi(t,u)$ converge to the stable
  equilibrium points, which form the lower boundary of an ellipse in the $(u,w)$-plane.}}\label{Fig:ellipsis}
\end{figure}

In the model of \citet{Heston}, the log-price $\Xt$ and the
corresponding variance process $\Vt$ are given under the
risk-neutral measure by the SDE
\begin{align*}
dX_t &= -\frac{V_t}{2}\,dt + \sqrt{V_t}\,dW^1_t\\
dV_t &= -\lambda(V_t - \theta)\,dt + \zeta \sqrt{V_t}\,dW^2_t\,
\end{align*}
where $W_t^1, W_t^2$ are Brownian motions with correlation
parameter $\rho$, and $\zeta, \lambda, \theta > 0$. In affine
form, the model is written as
\begin{subequations}\label{Eq:FR_Heston}
\begin{align}
F(u,w) &= \lambda \theta w\\
R(u,w) &= \frac{1}{2}(u^2 - u) + \frac{\zeta^2}{2}w^2 - \lambda w
+ uw \rho \zeta\;.
\end{align}
\end{subequations}
It is easily calculated that $\chi$ is given by $\chi(u) = \rho
\zeta u - \lambda$. We will first analyze the long term behavior
of $\Xt$, with the help of Theorem~\ref{Thm:wm_convergence}. To
satisfy the condition $\chi(1) < 0$ we need $\lambda > \zeta
\rho$. Note that this condition is always satisfied if $\rho \le
0$, the case that is typical for applications. Solving a quadratic
equation we find that
\[w(u) = \frac{(\lambda  - u \rho \zeta) - \sqrt{(\lambda - u \rho \zeta)^2 - \zeta^2 (u^2 - u)}}{\zeta^2}\;, \quad \text{and} \quad h(u) = \lambda \theta w(u)\;.\]
Denoting the term under the square root by $\Delta(u)$, we see
that $w(u)$ and $h(u)$ are both defined on $J = I = \set{u:
\Delta(u) \ge 0}$. Since $R$ is a second order polynomial in the
Heston model, the equilibrium points of the generalized Riccati
equation for $\psi$ form an ellipse in the $(u,w)$-plane, and
$w(u)$ is given by its lower part -- see Figure~\ref{Fig:ellipsis}
for an illustration. Interestingly, $w(u)$, and also $h(u)$, are
cumulant generating functions of a Normal Inverse Gaussian
distribution (cf. \citet[Eq.~(2.4)]{BarndorffNIG}). Thus, for
large $t$, the price process of the Heston model is, in terms of
its marginal distributions, close to a
Normal-Inverse-Gaussian exponential-L{\'{e}}vy model.\\

Next we consider moment explosions in the Heston model. As
mentioned above, moment explosions in the Heston model (and other
models) have already been studied by \citet{Piterbarg}.
Nevertheless this will provide a first test of
Theorem~\ref{Thm:moment_explosions}: In the case of the Heston
model it is easily determined from \eqref{Eq:FR_Heston} that
$f_+(u) = r_+(u) = \infty$. Calculating the integral in case
\eqref{Eq:moments_sub2} of Theorem~\ref{Thm:moment_explosions}, we
obtain
\begin{equation}\label{Eq:Heston_tstar}
T_*(u) =
    \begin{cases}
        +\infty \quad & \Delta(u) \ge 0\\
        %\frac{1}{\sqrt{\Delta}}\log\left(\frac{b + \sqrt{\Delta}}{b - \sqrt{\Delta}}\right) \quad &b>0, \Delta > 0\\
        \frac{2}{\sqrt{-\Delta(u)}} \left(\arctan \frac{\sqrt{-\Delta(u)}}{\chi(u)} + \pi \Ind{\chi(u) < 0}\right) \quad & \Delta(u) < 0\;.
    \end{cases}
\end{equation}
In Figure~\ref{Fig:moment_ex} a plot of this function for typical
parameter values is shown. Note that \cite{Piterbarg} distinguish
an additional case where $\chi(u) < 0$, but $\Delta(u)
> 0$. A little calculation shows that this can only happen if
$\chi(1) \ge 0$, a case that is precluded by our assumptions in
Theorem~\ref{Thm:wm_convergence}, and never occurs when $\rho \le 0$.\\

We will now study the effect of adding jumps to the Heston model.
The simplest case is the addition of an independent jump component
with constant activity: Let $\Jt$ be a pure-jump L\'evy process,
independent of $(W^{1,2}_t)_{t \ge 0}$ and define the
Heston-with-jumps model by
\begin{align*}
dX_t &= \left(\delta - \frac{V_t}{2}\right)\,dt + \sqrt{V_t}\,dW^1_t + dJ_t\\
dV_t &= -\lambda(V_t - \theta)\,dt + \zeta\sqrt{V_t}\,dW^2_t\;.
\end{align*}
The drift $\delta$ is determined by the martingale condition for
$\St$. To make the example simpler, we assume that $\Jt$ jumps
only downwards. This is equivalent to saying that the L\'evy
measure $m(dx)$ of $\Jt$ is supported on $(-\infty,0)$. The affine
form of the model is
\begin{align}
F(u,w) &= \lambda \theta w + \widetilde{\kappa}(u)\\
R(u,w) &= \frac{1}{2}(u^2 - u) + \frac{\zeta^2}{2}w^2 - \lambda w
+ uw \rho \zeta\;,
\end{align}
where $\widetilde{\kappa}(u)$ is the compensated cumulant
generating function of the jump part, i.e.
\[\widetilde{\kappa}(u) = \int_{(-\infty,0)}\left(e^{xu} - 1\right)\,m(dx) - u \int_{(-\infty,0)} \left(e^x - 1\right)\,m(dx)\;.\]
Let $\kappa_- < 0$ be the number such that $\widetilde{\kappa}(u)$
is finite on $(\kappa_-,\infty)$ and infinite outside.
For example, if the absolute jump heights are exponentially distributed with an expected jump size of $1/\alpha$, then $\kappa_- = - \alpha$.\\
To analyze the explosion times of this model, note that $R$, and
thus $\chi(u)$, $w(u)$, $I$ and $r_+(u)$ have not changed compared
to the Heston model. As long as $u > \kappa_-$, the explosion time
$T_*(u)$ is the same as in the Heston model. However, if $u \le
\kappa_-$, $F(u,0) = \infty$ and by
Theorem~\ref{Thm:moment_explosions}, $T_*(u) = 0$. Thus, the
addition of jumps to the Heston model has the effect of truncating
the explosion time to
zero, whenever $u \le \kappa_-$.\\
From the viewpoint of the critical moment functions, $u_+(t)$ does
not change compared to the Heston model, but $u_-(t)$ does; in the
model with jumps it is given by
\[u_-^\text{Jump}(t) = u_-^\text{Heston}(t) \vee \kappa_-\;.\]
Since $u_-$ is increasing with $t$, it makes sense to define a
cutoff time $T_\sharp$ by
\[T_\sharp = \sup \set{t \ge 0: u_-^\text{Heston}(t) = \kappa_-} = T_*(\kappa_-)\;,\]
such that
\begin{align*}
u_-^\text{Heston}(t) < u_-^\text{Jump}(t), &\quad \text{if} \quad t < T_\sharp\\
u_-^\text{Heston}(t) = u_-^\text{Jump}(t), &\quad \text{if} \quad
t \ge T_\sharp\;.
\end{align*}
In Figure~\ref{Fig:moment_ex} a comparison of the critical moment
functions in the Heston model with and without jumps can be seen.
By Lee's moment formula, the critical moment $u_-(t)$ moving
closer to $0$ will cause the left side of the implied volatility
smile to become steeper. Thus the net effect of adding the jump
component $\Jt$ to the Heston model, is a steepening of the left
side of the smile for maturities smaller than $T_\sharp$. For
times larger than $T_\sharp$, the asymptotic behavior of the smile
(in the sense of Lee's formula) is \emph{exactly} the same as in
the Heston model without jumps. This corresponds well to the
frequently made observation (see e.g. \citet[Chapter~5]{Gatheral})
that a Heston model with jumps can be fitted well by first fitting
a (jump-free) Heston model to long maturities, and then
calibrating only the additional parameters to the full smile. In
fact \citeauthor{Gatheral} proposes (on heuristical grounds) the
concept of a `critical time' $T$, after which the influence of an
independent jump component on the implied volatility smile can be
neglected. The analysis of the Heston model with jumps is of
course easily extended to the case that $\Jt$ is not one-sided. In
that case the effects discussed above will be seen to affect also
the right side of the implied volatility smile.

\subsection{A model of Bates}
We consider now the model given by
\begin{align*}
dX_t &= \left(\delta - \frac{V_t}{2}\right)\,dt + \sqrt{V_t}\,dW^1_t + \int_D{x\,\widetilde{N}(V_t,dt,dx)}\\
dV_t &= -\lambda(V_t - \theta)\,dt + \zeta\sqrt{V_t}\,dW^2_t\;.
\end{align*}
where as before $\lambda, \theta, \zeta > 0$ and the Brownian
motions are correlated with correlation $\rho$. The jump component
is given by $\widetilde{N}(V_t,dt,dx) = N(V_t,dt,dx) -
n(V_t,dt,dx)$, where $N(V_t,dt,dx)$ is a Poisson random measure,
and its compensator $n(V_t,dt,dx)$ is of the
\emph{state-dependent} form $V_t \mu(dx) dt$, with $\mu(dx)$ the
L{\'e}vy measure given in \eqref{Eq:FR_form}. A model of this kind
has been proposed by \citet{Bates} to explain the time-variation
of jump-risk implicit in observed option prices.
\citeauthor{Bates} also proposes a second variance factor, which
we omit in this example, in order to remain in the scope of
Definition~\ref{Def:ASVM}. It would however not be difficult to
extend our approach to the two-factor Bates model, since the two
proposed variance-factors are mutually independent, causing the
corresponding generalized Riccati equations to decouple. Since it
is affine, the above model can be characterized in terms of the
functions $F$ and $R$:
\begin{align}
F(u,w) &= \lambda \theta w\\
R(u,w) &= \frac{1}{2}(u^2 - u) + \frac{\zeta^2}{2}w^2 - \lambda w
+ uw \rho \zeta + \widetilde{\kappa}(u)\;.
\end{align}
where $\widetilde{\kappa}(u)$ is the compensated cumulant
generating function of the L{\'e}vy measure $\mu$. As in the
Heston model we can obtain $w(u)$ and $h(u)$ explicitly, and get
\[h(u) = \frac{-\chi(u) - \sqrt{\Delta(u)}}{\zeta^2}\;, \quad \text{and} \quad h(u) = \lambda \theta w(u)\;,\]
where $\chi(u) = \rho \zeta u - \lambda$ and $\Delta(u) =
\chi(u)^2 - \zeta^2 (u^2 - u + 2 \widetilde{\kappa}(u))$. Both
$w(u)$ and $h(u)$ are defined on $I = J = \set{u: \Delta(u) \ge
0}$. The time of moment explosion can again be calculated
explicitly, and is given by
\begin{equation}\label{Eq:Bates_tstar}
T_*(u) =
    \begin{cases}
        +\infty \quad & \Delta(u) > 0\\
        %\frac{1}{\sqrt{\Delta}}\log\left(\frac{b + \sqrt{\Delta}}{b - \sqrt{\Delta}}\right) \quad &b>0, \Delta > 0\\
        \frac{2}{\sqrt{-\Delta(u)}} \left(\arctan \frac{\sqrt{-\Delta(u)}}{\chi(u)} + \pi \Ind{\chi(u) < 0}\right) \quad & -\infty < \Delta(u) < 0\\
        0 \quad & \Delta(u) = -\infty\;.
    \end{cases}
\end{equation}

\subsection{The Barndorff-Nielsen-Shephard model} \label{Sec:BNS}

The Barndorff-Nielsen-Shephard (BNS) model was introduced by
\citet{BNS} as a model for asset pricing. In SDE form it is given
in the risk-neutral case by
\begin{align*}
    dX_t &= (\delta - \frac{1}{2} V_t)dt + \sqrt{V_t}\,dW_t +
    \rho\,dJ_{\lambda t}\\
    dV_t &= -\lambda V_t \,dt + dJ_{\lambda t}
\end{align*}
where $\lambda > 0$, $\rho < 0$  and $(J_t)_{t \geq 0}$ is a
L\'evy subordinator, i.e. a pure jump L\'evy process that
increases a.s. The drift $\delta$ is determined by the martingale
condition for $\St$. The time-scaling $J_{\lambda t}$ is
introduced by \citeauthor{BNS} to make the invariant distribution
of the variance process independent of $\lambda$. The distinctive
features of the BNS model are that the variance process has no
diffusion component, i.e. moves purely by jumps and that the
negative correlation between variance and price movements is
achieved by simultaneous jumps in $\Vt$ and $\Xt$. The BNS model
is an affine stochastic volatility model, and $F$ and $R$ are
given by
\begin{align}
F(u,w) &= \lambda \kappa(w + \rho u) - u \lambda \kappa(\rho)\\
R(u,w) &= \frac{1}{2}(u^2 - u) - \lambda w
\end{align}
where $\kappa(u)$ is the cumulant
generating function of $(J_t)_{t \geq 0}$.\\
We simply have $\chi(u) = - \lambda$ and $w(u)$ from
Lemma~\ref{Lem:existence} is given by
\[w(u) = \frac{1}{2\lambda}(u^2 - u)\;.\]
It follows that
\[h(u) = \lambda \kappa \left(\frac{u^2}{2\lambda} + u\left(\rho - \frac{1}{2\lambda}\right)\right) - u\lambda\kappa(\rho)\;.\]
This expression can be interpreted as cumulant generating function
of a Brownian motion with variance $\frac{1}{\lambda}$ and drift
$\rho - \frac{1}{2\lambda}$, subordinated by the L\'evy process
$J_{\lambda t}$ and then mean-corrected to satisfy the martingale
condition.\\
To analyze moment explosions in the BNS model, let $\kappa_+ :=
\sup \set{u
> 0 : \kappa(u) < \infty}$. It is easy to see that $f_+$ is given by $f_+ = \max(\kappa_+ -
\rho u,0)$. Since $r_+  = \infty$, we have that the explosion time
for the moment of order $u$ is given by
\[T_*(u) = \int_0^{f_+}\frac{d\eta}{R(u, \eta)} = -\frac{1}{\lambda} \log \left(1 - \frac{2\lambda (\max(\kappa_+ - \rho u,0))}{u(u - 1)}\right)\;.\]
The critical moment functions $u_\pm(T)$ can be obtained
explicitly by solving a quadratic equation, and are given by
\[u_\pm(t) = \frac{1}{2} - \frac{\rho \lambda}{1 - e^{-\lambda t}} \pm \sqrt{\frac{1}{4} + \frac{(2 \kappa_+ - \rho) \lambda}{1 - e^{-\lambda t}} + \frac{\rho^2 \lambda^2}{\left(1 - e^{-\lambda t}\right)^2}}\;.\]
The large-strike asymptotics for the implied volatility smile in
the sense of Lee can be explicitly calculated by inserting $u_\pm$
into Proposition~\ref{Prop:Lee}.

\subsection{The Heston model in the stationary variance regime}
In the Heston model the limit distribution of the variance process
$\Vt$ is a Gamma distribution with parameters $(-\frac{2 \lambda
\theta}{\zeta^2}, \frac{2 \lambda}{\zeta^2})$. This is well-known,
but can also be obtained by applying
Proposition~\ref{Prop:stationary}. The cumulant generating
function $l(w)$ is thus given by
\[l(w) = -\frac{2 \lambda \theta}{\zeta^2} \log \left(1 - \frac{\zeta^2}{2 \lambda} w\right)\;,\]
defined on $(-\infty,\frac{2 \lambda}{\zeta^2})$, such that $l_+ =
\frac{2 \lambda}{\zeta^2}$. As before we have that $\chi(u) = \rho
\zeta u - \lambda$, and we assume that $\chi(1) < 0$. In addition
we define $\chi^+(u) = \rho \zeta u + \lambda$. By
Theorem~\ref{Thm:moment_explosions_stationary}, the explosion time
in the stationary regime is given by
\begin{multline}\label{Eq:Heston_tstar_stationary}
T_*^S(u) = \int_0^{2\lambda / \zeta^2}{\frac{d\eta}{R(u, \eta)}} = \\
= \begin{cases}\infty \quad &\sqrt{\Delta(u)}  > - \chi^+(u) , \\[7pt]
\frac{1}{\sqrt{\Delta}}\log\left|\frac{\chi^+ \chi + 2 \lambda \sqrt{\Delta} - \Delta}{\chi^+ \chi - 2 \lambda \sqrt{\Delta} - \Delta}\right| \quad &0 < \sqrt{\Delta(u)}  < - \chi^+(u),\\[7pt]
 \frac{2}{\sqrt{-\Delta}} \arctan  \left(\frac{2 \lambda \sqrt{-\Delta}}{\chi^+ \chi - \Delta}  + \pi \Ind{\chi^+ \chi < \Delta}\right) \quad & \Delta(u) < 0\;.\end{cases}
\end{multline}

\begin{figure}[tbp]
  \includegraphics[height=.45\textheight]{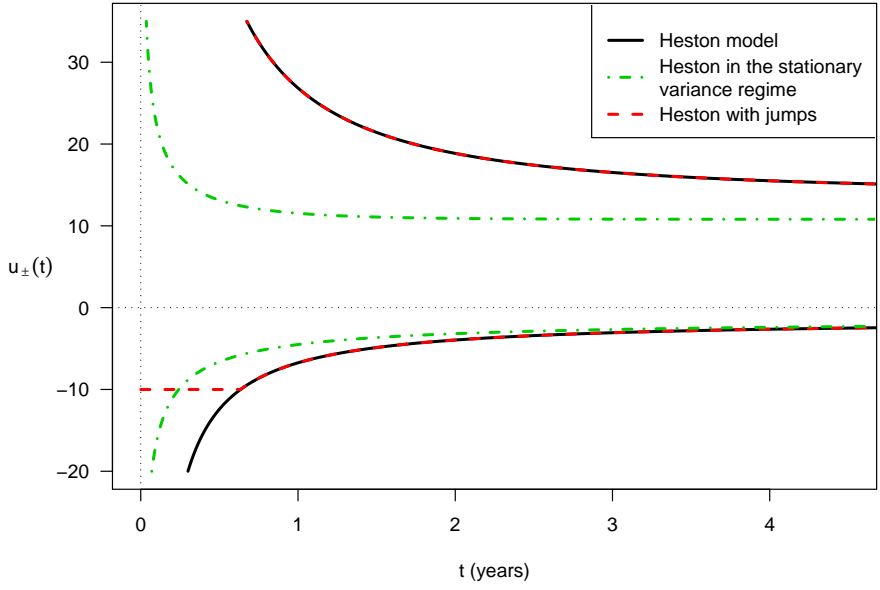}\\[-5pt]
  \caption{{\Small This plot shows the critical moment functions $u_\pm(t)$ for a Heston model with the same parameters as in Figure~\ref{Fig:ellipsis}.
  Also shown are $u^S_\pm(t)$ for the model in the stationary variance regime, and $u^\text{Jump}_\pm(t)$
  for the Heston model with an independent jump component, whose negative jump heights are exponentially distributed with mean $\alpha = -0.1$. Note that $u^\text{Jump}_\pm(t)$
  coincides with $u_\pm(t)$ everywhere except in the lower left corner of the plot.}}
  \label{Fig:moment_ex}
\end{figure}

In Figure~\ref{Fig:moment_ex} $T_*^S(u)$ is plotted together with
$T_*(u)$ for the Heston model.

\subsection{The BNS model in the stationary variance regime}

In the BNS model, the cumulant generating function of the limit
distribution $L$ of the variance process is given by
Proposition~\ref{Prop:stationary} by
\[l(w) = \int_0^w \frac{\kappa(\eta)}{\eta} d\eta\;,\]
provided the log-moment condition $\int_{y > 1}(\log y)\, \mu(dy)
< \infty$ holds for the L\'evy measure of $(J_t)_{t \geq 0}$. The
above integral is finite as long as $w \in (-\infty, \kappa_+)$,
and infinite outside. Thus $l_+ = \kappa_+$. In
Section~\ref{Sec:BNS} we obtained that $f_+(u) = \kappa_+ - \rho
u$, such that the time of moment explosion under stationary
variance is given by
\[T_*^S(u) = \int_0^{min(f_+(u),l_+)} \frac{d\eta}{R(u, \eta)} = -\frac{1}{\lambda} \log \left(1 - \frac{2\lambda k(u)}{u(u - 1)}\right)\;,\]
where $k(u) = \kappa_+$ for $u \ge 1$ and $k(u) = \max(\kappa_+ -
\rho u,0)$ for $u \le 0$. Again, this expression can be inverted
to give the critical moment functions in the stationary variance
case. By definition $\rho \le 0$, such that we obtain
\begin{align*}
u_-^S(T) &= \frac{1}{2} - \frac{ \rho \lambda}{1 - e^{-\lambda T}} - \sqrt{\frac{1}{4} + \frac{(2 \kappa_+ - \rho)\lambda}{1 - e^{-\lambda T}} + \frac{\rho^2  \lambda^2}{\left(1 - e^{-\lambda T}\right)^2}}\\
u_+^S(T) &= \frac{1}{2} + \sqrt{\frac{1}{4} + \frac{2 \kappa_+
\lambda}{1 - e^{-\lambda T}}} \;.
\end{align*}

\appendix

\section{Additional proofs}

\begin{proof}[Proof of Theorem~\ref{Thm:Riccati}]
Let $t \le \tau$. By the flow equation we can write
\begin{align*}
\phi(\tau,u,\eta) &= \phi(t,u,\eta) + \phi(\tau - t,u,\psi(t,u,\eta))\\
\psi(\tau,u,\eta) &= \psi(\tau-t,u,\psi(t,u,\eta))\;.
\end{align*}
Since the left sides are finite by assumption, it follows that
also $\phi(t,u,\eta)$ and $\psi(t,u,\eta)$ are. $V_t$ is
non-negative, such that
\[\left|\EE\left[\exp\left(u X_t + w V_t \right)\right]\right| \le \left|\EE\left[\exp\left(u X_t + \eta V_t \right)\right]\right|\;,\]
whenever $\Re\,w \le \Re\,\eta$. Thus $\phi(t,u,w)$ and
$\psi(t,u,w)$ exist for all $w \in \CC$ with $\Re\,w \le
\Re\,\eta$. As a particular case we can conclude that
$\phi(t,u,w)$ and $\psi(t,u,w)$ exist for all $(u,w)$ in $\cU :=
\set{(u,w) \in \CC^2: \Re\,u = 0, \Re\,w \le 0}$.\\
We also define $\cU^\circ := \set{(u,w) \in \CC^2: \Re\,u = 0,
\Re\,w < 0}$, and show next that $\phi(t,u,w)$ and $\psi(t,u,w)$
are (right-)differentiable at $t = 0$ for all $(u,w) \in
\cU^\circ$. The key idea of our proof is originally due to
\citet{MontgomeryZippin}, and has also been presented in
\citet{FilipovicTeichmann} and \citet{DawsonLi}. First note that
the identity
\[\EE\left[w V_t e^{u X_t + w V_t}\right] = \left(\pd{}{w}\phi(t,u,w) + V_0 \pd{}{w}\psi(t,u,w)\right) \exp\left(\phi(t,u,w) + V_0 \psi(t,u,w) + X_0 u\right)\]
shows that $\pd{}{w}\phi(t,u,w)$ and $\pd{}{w}\psi(t,u,w)$ exist,
and are continuous for all $t \le \tau$ and $(u,w) \in \cU^\circ$.
By Taylor expansion it holds that
\begin{align}\label{Eq:Phi_interm1}
\int_0^s{\psi(r,u,\psi(t,u,w))\,dr} - \int_0^s{\psi(r,u,w)\,dr} &=
\int_0^s{\pd{}{w}\,\psi(r,u,w)\,dr} \left(\psi(t,u,w)-w\right)
\notag \\ &+ o\big(|\psi(t,u,w) - w|\big)\;.
 \end{align}
On the other side, using the flow property, we calculate
\begin{align}\label{Eq:Phi_interm2}
&\int_0^s{\psi(r,u,\psi(t,u,w))\,dr} - \int_0^s{\psi(r,u,w)\,dr} =
\int_0^s{\psi(r + t,u,w)\,dr} - \int_0^s{\psi(r,u,w)\,dr} = \notag \\
= &\int_t^{s + t}{\psi(r,u,w)\,dr} - \int_0^s{\psi(r,u,w)\,dr} =
% \int_s^{s + t}{\psi(r,w)\,dr} - \int_0^t{\psi(r,w)\,dr} = \notag \\
 \int_0^t{\psi(r + s,u,w)\,dr} - \int_0^t{\psi(r,u,w)\,dr}\;.
\end{align}
Denoting the last expression by $I(s,t)$, and putting
\eqref{Eq:Phi_interm1} and \eqref{Eq:Phi_interm2} together, we
obtain
\[\lim_{t \to 0}\frac{\left|\frac{1}{s}I(s,t)\right|}{\left|\psi(t,u,w) - w\right|} =
\left|\frac{1}{s}\int_0^s{\pd{}{w}\psi(t,u,w)\,dr}\right|\;.\]
Thus, writing $M_s = \frac{1}{s}
\int_0^s{\pd{}{w}\psi(t,u,w)\,dr}$, we have
\[\lim_{t \to 0}\frac{1}{t}\left|\psi(t,u,w) - w\right| = \left|\lim_{t \to 0} \frac{I(s,t)}{st}\right| \cdot \left|M_s\right|^{-1} = \left|\frac{\psi(s,u,w) - w}{s}\right| \left|M_s\right|^{-1}\;.\]
But $M_s$ is a continuous function of $s$, and $\lim_{s \to 0} M_s
= \pd{}{w} \psi(0,u,w) = 1$, such that for $s$ small enough $M_s
\neq 0$. We conclude that the left hand side is finite, and using
\eqref{Eq:Phi_interm1} we obtain that
\[\lim_{t \to 0} \frac{\psi(t,u,w) - w}{t} = \left(\frac{\psi(s,u,w) - w}{s}\right)  \cdot \left(\frac{1}{s}\int_0^s{\pd{}{w}\psi(r,u,w)\,dr}\right)^{-1}\;.\]
The finiteness of the right hand side implies the existence of the
limit on the left. In addition the right hand side is continuous
for $(u,w) \in \cU^\circ$, showing that also the left hand side
is. A similar calculation for $\phi(t,u,w)$ shows that
\[\lim_{t \to 0} \frac{\phi(t,u,w)}{t} = \frac{\phi(s,u,w)}{s} - \lim_{t \to 0}\left(\frac{\psi(t,u,w) - w}{t}\right)  \cdot \left(\frac{1}{s}\int_0^s{\pd{}{w}\phi(r,u,w)\,dr}\right)\;,\]
allowing the same conclusions for $\phi(t,u,w)$. We have thus
shown that the time-derivatives of $\phi(t,u,w)$ and $\psi(t,u,w)$
at $t = 0$ exist, and are continuous in $\cU^\circ$. Combining
\citet[Proposition~7.2]{Schachermayer} and
\citet[Proposition~6.4]{Schachermayer} the differentiability can
be extended from $\cU^\circ$ to $\cU$, and we have shown that
$(X_t,V_t)_{t \geq 0}$ is a \emph{regular} affine process. The
rest of Theorem~\ref{Thm:Riccati} follows now as in
\citet[Theorem~2.7]{Schachermayer}
\end{proof}

\begin{proof}[Proof of Lemma~\ref{Lem:psiphi_prop}]
We prove the assertions of Lemma~\ref{Lem:psiphi_prop} for $F$;
they follow analogously for $R$. By the L\'evy-Khintchine
representation \eqref{Eq:FR_form}, $F(u,w) + c$ is the cumulant
generating functions of some infinitely divisible random
variables, say $X$. Writing $z = (u,w) \in \RR^2$, and using
H\"older's inequality it holds for any $\lambda \in [0,1]$ that
\begin{multline}\label{Eq:F_hoelder}
F(\lambda z_1 + (1-\lambda) z_2) = \log \EE\left[e^{\lambda
\scal{z_1}{X}}e^{(1-\lambda) \scal{z_2}{X}}\right] - c\le \\
\le \lambda \log \EE\left[e^{\scal{z_1}{X}}\right] + (1 - \lambda)
\EE\left[e^{\scal{z_2}{X}}\right] - c = \lambda F(z_1) +
(1-\lambda) F(z_2)\;,
\end{multline}
showing convexity of $F$. In addition equality in
\eqref{Eq:F_hoelder} holds if and only if $k e^{\scal{z_1}{X}} =
e^{\scal{z_2}{X}}$ a.s. for some $k
> 0$. This in turn is equivalent to $\scal{z_1 - z_2}{X}$
being constant a.s. Choosing now $z_1$ and $z_2 \neq z_1$ from
some one-dimensional affine subspace $U = \set{p + \scal{q}{x} : x
\in \RR}$ of $\RR^2$, we see that either $\scal{q}{X}$ is constant
a.s. in which case $F|_U$ is affine, or it is not constant, in
which case strict inequality holds in \eqref{Eq:F_hoelder} for all
$z_1, z_2 \in U$,
showing (c).\\
Let $L_\alpha = \set{z: F(z) \le \alpha}$ be a level set of $F$,
and $z_n \in L_\alpha$ a sequence converging to $z$. Then by
Fatou's Lemma
\[\log \EE[e^{\scal{z}{X}}] - c \le \liminf_{n \to \infty} \log \EE[e^{\scal{z_n}{X}}] - c \le \alpha\;,\]
showing that $z \in L_\alpha$ and thus that $F$ is a closed convex
function. Finally $F$ is proper, because $F(0,0) = c > -\infty$,
showing (a).\\
Next we show analyticity: Consider the random variables $X_n := X
\Ind{|X| \le n}$. Since they are bounded, their Laplace
transforms, and hence also their cumulant generating functions are
entire functions on $\CC^2$, and thus analytic on $\RR^2$. As a
uniform limit of analytic functions $F(u,w)$ is analytic in the
interior of $\dom\,F$, showing (b). Assertion (d) follows directly
from Theorem~\ref{Thm:Riccati}.
\end{proof}

%
%
%
%
%-------------------------------
\bibliographystyle{plainnat}
\bibliography{affine_smile}
%-------------------------------
%
%
%
\end{document}